\documentstyle[prl,aps,epsf,psfig,multicol]{revtex}
\begin{document}
\draft 
\title{The $W$ center in self-implanted silicon
is the   self-interstitial cluster I$_3$}
\author{Giorgia M. Lopez and Vincenzo  Fiorentini}
\address{INFM and Dipartimento di  Fisica, Universit\`a di Cagliari,
Cittadella  Universitaria, I-09042 Monserrato (CA), Italy}
\date{\today} 
\maketitle

\begin{abstract}
We identify the $W$ center in
self-implanted crystalline Si 
 with the three-membered self-interstitial
 cluster I$_3$ on the basis of
 first-principles density-functional-theory calculations  matching
  all the known experimental signatures of the center 
 (emission energy, extrinsic energy levels, 
activation energy and dissociation energy, local vibrational 
structure, and symmetry).
\end{abstract}

\begin{multicols}{2}
Self-implanted Si is relevant to the pre-amorphization processes
used at various stages of device fabrication, and as a model
environment for the study of anomalous interstitial-driven impurity
diffusion.
A prominent defect in self-implanted Si is 
 the so-called $W$ center, which has been known and studied 
intensely for a long time  \cite{old,davies}, but not firmly
identified so far, although generally accepted to
 originate  from the clustering of excess interstitial Si
atoms \cite{coffa1,coffa2,solosi}. By means of first-principles
density-functional calculations, we provide a conclusive 
identification of the $W$ center with the I$_3$ 
three-membered 
self-interstitial cluster. Remarkably for a difficult enterprise
such as the  identification of defects in crystals, our results 
 match  all the  experimental signatures 
available for this center,
and confirm the relevance of ab initio calculations for
the  physics of defects and impurities in crystals.

Recent experimental studies \cite{coffa1,coffa2}, 
performed with a variety of combined techniques, including variable-dose
implant, rapid thermal annealing (RTA), deep  level transient
spectroscopy (DLTS), and photoluminescence (PL), have firmly
identified   several  experimental signatures of the $W$ defect.
By carefully cross-referencing a number of  $n$-type samples
with different implant and RTA histories, Libertino {\it et al.}
\cite{coffa1} identified a low-dose implant regime in which the extended
\{311\} defects -- a common product of self-interstitial aggregation --
 do  not form:  small, nearly  point-like entities
are active instead. They suggested that  the  $W$ photoluminescence
line, a typical emission at 1.018 eV in self-implanted Si,   
is indeed the main signature of these centers. A second signature 
 identified as strongly correlated   with this emission, is a pair of
extrinsic levels determined by DLTS at about 0.35 and 0.6 eV above the
valence edge. The third  signature brought forth by the
 analysis of the pertaining DLTS signals under thermal treatment is the
activation energy of around 2.3 eV needed to cause the essentially
concurrent disappearance of these two levels upon RTA.
In addition, the appearance
of the $W$ band is itself thermally activated, with a characteristic
energy of about 0.85 eV \cite{coffa2,act-en} -- the fourth signature
 of $W$. A fifth group of signatures stems from the vibronic
structure of which  the $W$ emission is the zero-phonon line; this
structure has been  known for some time \cite{davies} to include main
peaks  at 17 meV and 40 meV, and a local vibrational mode at 70 meV 
\cite{davies}. The symmetry of the center identified via 
the stress response of the  vibronic structure is at least 
 C$_{3v}$. Finally, several indirect hints led many authors (see e.g.
 \cite{coffa1} for a summary)  to postulate the most likely 
size  of the cluster responsible for these signatures to be $n$=3, 
with a high-symmetry compact structure, as opposed to
the  non-compact I$_4$ and asymmetric I$_2$ 
clusters  found in recent calculations
 \cite{noi1,bongiorno,kim0}. 

In the following, we give evidence that the I$_3$ Si  tri-interstitial is 
to be identified  with the $W$ center, as its properties explain all 
the mentioned  experimental signatures.
Our conclusion is based on a series of ab initio calculations on
self-interstitial  clusters in $c$-Si (full details are reported
elsewhere\cite{noi1}), performed at the 
first-principles level  within density-functional theory in the generalized
gradient approximation (GGA) \cite{pw91}, and the ultrasoft 
pseudopotential-plane
wave repeated-supercell approach using the Vienna Ab-initio 
Simulation Package code \cite{vasp}. We report data obtained 
with 32- and  64-atom supercells (the results are in fact
quite insensitive to cell size), multiprojector 
ultrasoft pseudopotentials \cite{vasp} (with two $s$, two $p$, and one $d$ 
projectors, r$_{c}$=1.31 \AA),
a plane wave cutoff of 151 eV, and a 444 k-space summation
 mesh (this  yields  32 k-points in the Brillouin zone
 for $C_1$,  i.e. no, imposed
symmetry; in occasional test we used 666 and 888 meshes, with 108 and 
256 k-points  
respectively). Atomic geometries are relaxed until 
all force components are below 0.01 eV/\AA. The theoretical lattice constant
$a_{\rm Si}$=5.461 \AA\, is used throughout. The formation energies 
and extrinsic levels are obtained in the usual way (see
e.g. Refs. \cite{eprime} and \cite{vdw}), including multipole
corrections for charged states \cite{multipole}. 

\begin{center}
\begin{figure}[h]
\begin{center}
\epsfclipon \epsfxsize=7cm \epsffile{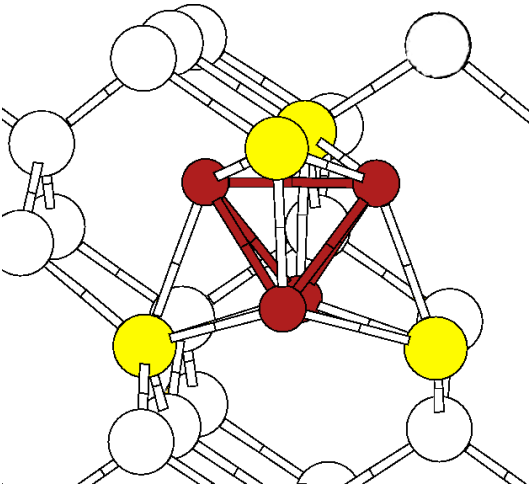}
\end{center}
\caption {View of the structure of the I$_3$ self-interstitial cluster
 in $c$-Si. Cluster atoms  are dark grey, while their
first  bulk-like neighbors,  involved in the 70-meV local vibrational
mode (see text),  are light grey. Bulk atoms are white.}  
\label{fig1}
\end{figure}
\end{center}

We obtained the equilibrium structure of the I$_3$ cluster by
 relaxation of a supercell containing an I$_2$ di-interstitial cluster
 \cite{noi1,bongiorno,kim0} and a single interstitial I$_1$ in the
 adjacent tetrahedral site, both in the neutral state. The
 interstitial binds spontaneously to I$_2$, producing the I$_3$
 structure depicted in Fig. \ref{fig1}. This structure agrees well
 with recent tight-binding \cite{bongiorno} and ab initio
 results\cite{kim}, although not with others \cite{coomer}
 (incidentally, the properties of the $n$=3 cluster studied in
 \cite{coomer} do not correlate well with those of the $W$ center).
The interatomic distances between the four atoms (dark grey in
 Fig. \ref{fig1}) involved in the local structure of I$_3$ are
 essentially identical, 2.488$\pm$0.002 \AA\, or 1.054 times the
 calculated Si-Si bond length of 2.365 \AA. All bonds are oriented
 along (110)-equivalent directions, hence the   four atoms form a
 perfect tetrahedron with edges aligned 
 with the cubic (110) axes, and whose centerpoint is at a lattice
 site. The local symmetry of the cluster is therefore T$_d$.  This
 agrees with the response of the zero-phonon $W$ line to applied
 stress and its dependence on the electric field direction. The
 symmetry of the cluster, we point out, is obtained spontaneously in
 the simulated assembly of I$_2$ and tetrahedral I$_1$ {\it without}
 imposing any initial symmetry.

Studying the energies of different charge states of the defect, we
 determined the extrinsic levels. I$_3$ possesses two extrinsic
 thermal charging levels in the gap, namely $\epsilon$(++/+)=0.35 eV +
 E$_v$ and $\epsilon$(+/0)=0.8 eV + E$_v$, with E$_v$ the valence band
 top. These values are close to the experimental values of 0.35 eV and
 0.6 eV.  Since the samples of interest here are $n$-type, these
 levels are filled and DLTS-detectable. Since a supersaturation of Si
 interstitials in $c$-Si produces moderate $n$-type conditions
 \cite{noi1}, this conclusion will hold even in as-implanted intrinsic
 samples. The DLTS signatures of the $W$ center are thus reproduced by
 the I$_3$ cluster.

To interpret the thermal evolution of these levels, we assume
that their disappearance under thermal treatment is due to the 
evaporation of the cluster into single interstitials. For the
 $n$-type Fermi level of relevance  here, we find that all the interstitial
complexes I$_1$, I$_2$, and I$_3$ are in their neutral state. We 
then estimate the cost of splitting I$_3$ into well-separated neutral
I$_2$ and (neutral dumbbell) I$_1$ by direct comparison of calculated
total energies :  this cost is 2.38 eV, matching very closely the
 deactivation energies, 2.28 eV and 2.36 eV (estimated experimental 
error $\simeq$ 15\%), of the observed DLTS peaks, confirming our
previous attribution of those peaks to the I$_3$ extrinsic levels. 

After splitting an I$_3$ into an I$_2$ and an I$_1$, one may still
have electrically active levels present, since both the latter centers
have extrinsic levels \cite{noi1} in the 0.25-0.35 eV and 0.6-0.8 eV
range. The observation of those levels is however preempted by two
facts: {\it a)} I$_2$ splits into two I$_1$'s with an energy cost of
1.5 eV, hence the thermal treatment dissolving the I$_3$'s also gets
rid of any I$_2$'s; {\it b)} an I$_3$-cracking thermal process should
presumably cause the remaining I$_1$'s (also electrically 
active \cite{noi1}) to  evaporate off the sample, both because they
are quite rapidly diffusing \cite{chang} neutral dumbbells, and
because they have no effective capture centers, such as vacancies,
available in sufficient concentration.  

Of course the main signature of the center at issue is the $W$
emission itself. The relevant emission energy is 1.02 eV; since
the material is $n$-type, the transition is probably
bound-to-free. Since the gap of Si is about 1.1 eV at the relevant
temperature, the involved level lays $\sim$80 meV below the conduction
edge. This does not imply, however, a delocalized state as
in shallow impurities, the very existence of a sharp zero-phonon line
in an indirect gap material indicating a localized
character.  We find the 0/-- level of I$_3$ at 1.10 eV above the
valence (k-points convergence was checked to better than $\pm$10 meV
up to the (888) mesh; no change was observed between 32-, 64- and
265-atom cells). This is clearly a very good candidate for the initial state  
of the emission: its energy is quite close to the observed transition
and, since emission is involved, the Franck-Condon principle dictates
that the emission energy coincides with the charging level in this
case. In addition, this state is empty in thermal equilibrium, and can 
thus receive the photoexcited electrons thermalizing down from the upper
conduction states.  We find  that the state is orbitally
non-degenerate, in agreement with the analysis of the phonon replicas  
 and with  the absence of pseudo--Jahn-Teller distortions \cite{davies}.
To give a proper estimate of the position of the  state with
respect to the conduction band, we calculated the fundamental gap of Si
using the expression, exact in the $N\rightarrow\infty$ limit
\cite{sham},
\begin{equation}  
$$E_{\rm gap} = E_{\rm tot}(N+1) - 2 E_{\rm tot}(N) + E_{\rm
tot}(N-1),
\label{dscf}
\end{equation}
using an undefected 64-atom supercell with N, N+1, and N--1 electrons.
 The gap energy thus obtained is 1.13 eV, which gives a 30
meV binding energy. 
The satifactory agreement with experiment further
confirms our identification \cite{nota}.

We now consider the activation barrier for the {\it creation} of the
center, which was quantified in 0.85 eV in Refs. \cite{coffa2,act-en}.  The
simplest interpretation is that the diffusive motion of the
interstitials must be activated to achieve clustering. It is natural
to presume that at least two processes are to be activated to form
I$_3$: the diffusion of I$_1$'s towards each other to form I$_2$, and
the diffusion of I$_1$ towards I$_2$ to form I$_3$. We estimated (see
also \cite{chang}) a minimum energy barrier for the
dumbbell-to-dumbbell diffusion of 0.2 eV via a tetrahedral site and
0.18 eV via an hexagonal site. This is much lower that the
experimental activation energy. We then hypothesized that I$_1$ be
subject to a local repulsion by its companion center (another I$_1$ or
the I$_2$), effectively increasing its diffusion barrier. To check
this idea we compare the energy of a dumbbell self-interstitial and a
tetrahedral I$_1$ in the same cell in adjacent sites -- i.e. the last
saddle-point configuration before I$_2$ formation \cite{noi1} -- with the
sum of the energies of an isolated dumbbell I$_1$ plus an isolated
tetrahedral I$_1$. The difference is an estimate of the local
repulsion (if any) between single interstitials.  We then calculate
the same difference for I$_2$ and tetrahedral I$_1$ in the same cell,
and isolated I$_2$ plus isolated tetrahedral I$_1$.  Again, the
difference is an estimate of the local repulsion (if any) between
I$_1$ and I$_2$. In both cases we obtain an effective repulsion, of
0.64 eV in the first case, and 0.53 eV in the second case (errors due
to finite-size  relaxation effects are at least one order of magnitude
smaller \cite{noi1}).  Therefore, an effective repulsion acts locally
between the precursors of I$_3$ (I$_1$'s and I$_2$)
 The largest of these two repulsion energies,
added to the normal diffusion barrier, provides an estimate of the
maximum effective barrier.  The result is 0.84 eV, in close agreement
with the experimental estimate of 0.85 eV. Thus, an effective local
repulsion between the component centers of the I$_3$ cluster explains
quantitatively the activation-energy signature of the $W$-band center.

We now analyze the vibrational modes of the center.  The zero-phonon 
$W$ line splits under (110)- and (111)-oriented, but not under
(100)-oriented stresses \cite{davies}. This is 
quite compatible with the symmetry
and orientation of the I$_3$ complex,  a
tetrahedron with (111)-oriented axes and (110)-oriented bonds.  The
analysis of the phonon replicas of Ref.\onlinecite{davies} suggests a
dominance of couplings to the Si bulk phonon continuum.  The main
features in the replica spectrum\cite{davies} are broad
structures associated with vibrational energies of 17 meV and 
40 meV. Further, sharper but much weaker lines exist
 in the optical-mode energy region. An additional peak
 at 70 meV below the zero-phonon line appears to involve
\cite{davies} a local vibrational mode (the highest vibrational
energy in Si bulk is $\hbar\omega^{\rm Si}_{\rm TO}(\Gamma)$=64 meV).

\begin{center}
\begin{figure}[h]
\begin{center}
\epsfclipon
\epsfxsize=8cm
\epsffile{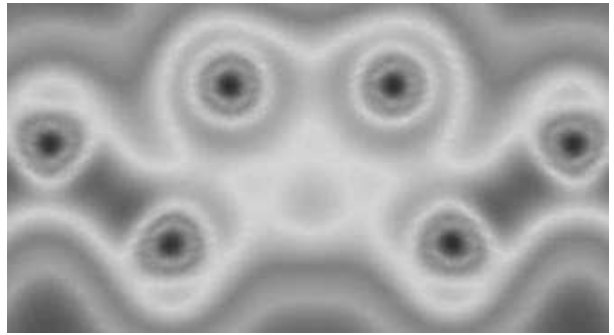}
\end{center}
\caption {Charge density contour plot in a (110) plane passing through
one of the internal bonds of the I$_3$ tetrahedron. Two of the cluster
atoms are visible (center top).  The 
amount of bonding charge between the cluster atoms, and towards 
the bulk atoms,  appears modest when compared to bulk-like bonds.}
\label{fig2}
\end{figure}
\end{center}

To identify possible local modes, we estimated the   vibrational
frequency of selected normal modes of  the cluster via the
frozen-phonon  method \cite{gp} in which a given   
displacement pattern is frozen into the lattice, and
the force/displacement ratio for the atoms involved  yields the 
mode's harmonic force constant. The calculated Si LO-TO mode energy 
at $\Gamma$ of 62.4 meV, --2.5\% from experiment\cite{gga-phon}
sets a reliability reference. The internal vibrational modes of the
cluster,  i.e. those involving the four atoms of the  cluster,
are found to have energies in the  range 
35--45 meV. The modes considered include most of 
the typical modes of a tetrahedron (breathing, twist, pinch, etc.)
 schematized e.g. in Fig. 2.5 of
  Ref. \onlinecite{vibronic}. A detailed
 discussion will be presented elsewhere. The relatively low
 frequencies of the internal modes  are unsurprising if we consider
the weaker bonding   within the cluster compared to the bulk: this is
apparent from  the (110)-plane slice of the charge density
\cite{vview} through one of the internal cluster bonds of neutral
I$_3$ in  Fig. \ref{fig2}. The ``translational'' mode of I$_3$, in
which the whole cluster moves with respect to the crystal, and which
would have zero frequency for the cluster in free space, has a 
low energy of 19 meV. This is expected since the cluster is also
quite  weakly  bound to the surrounding bulk (Fig. \ref{fig2}).

Their highest frequency being less than 45 meV, none of
 the internal vibrations can explain the 70-meV 
local mode \cite{davies}. Remarkably, indeed, it turns out that
 the local mode is not a proper cluster mode: 
 after some search we  identified it with the  vibration of 
each of the four cluster-adjacent atoms
 (light grey in Fig. \ref{fig1}) 
in the (111) planes parallel to the
corresponding  triangular face of the tetrahedral
cluster.  The mode  energy of 71.4 meV matches 
well the  observed 70 meV shift of the 
local-mode  replica. Its  isotropic
character (four equivalent atoms are involved, near each cluster
face), and its strong expected
 dependence on (110)- and (111)-oriented strains also agree with
the observed properties of the local-mode replica \cite{davies}. 
The mode clearly derives from a TO wave impinging on the I$_3$ 
cluster from a (111) direction -- pictorially,
  a ``breaker-on-the-shoal'' effect.

As for the prominent 17 meV and 40 meV replica structures, they can
be interpreted as suggested \cite{davies} as the interaction signature
of I$_3$ with the bulk phonon continuum, with some contribution, we add,
of the internal modes of the cluster. The phonon dispersion
and density of states  of Si \cite{gian} suggest that the modes
 involved in the 5-meV--wide structure at about 17 meV
 are mainly the zone-border modes at the K and X points (e.g. the
$\Sigma_3$ TA mode at K, e.g., has the correct 18 meV energy). The
``translation'' mode of the cluster mentioned earlier, with its energy
of 19 meV, is also expected to be involved in  this replica. For  the
40-meV feature,   the  phonon  density-of-states        weight comes
mainly  from the  LA modes   along the $\Lambda$ line,
topping   with  the 46-meV $L_2$ mode at L.  As mentioned,  the 35--45
meV  internal  modes of the clusters t are  also most likely to be
involved in the 40-meV replica, superposed on the bulk continuum, and
contributing to its substantial (10 meV) linewidth. With  these
attributions of vibrational signatures, in particular of the local
mode,   we  conclude our identification of the I$_3$  cluster with the
$W$ center.       

In summary, ab initio calculations of emission, activation, and
dissociation energy, extrinsic levels, and vibrational modes explain
all the known experimental  features associated with the $W$ band in
self-implanted Si (PL emission, DLTS signatures of electronic levels,
 RTA behavior, phonon replicas in the PL spectra) as being due  to  
the tri-interstitial cluster I$_3$. 
The W center remains thus unambiguously identified with the
I$_3$ self-interstitial cluster.

We acknowledge partial support from the Italian Ministry
of Research within the PRIN 2000 project ``Non-equilibrium dopant
diffusion in Si'', and from the Parallel Supercomputing Initiative of
INFM.

\end{multicols}
\end{document}